%
%
%
%
%
%
\input psfig.sty
%
%
\magnification=\magstep1
%
%
%
%
%
\parskip=2pt plus1pt
\parindent=2em
\baselineskip=13pt

%
%
\catcode`\@=11
\topskip=10pt plus 10pt
\r@ggedbottomtrue
\catcode`\@=12
%
%
\font\sslarge=cmss17
\font\allcaps=cmcsc10
%
%
\font\eightrm=cmr8
\font\sixrm=cmr6

\font\eighti=cmmi8
\font\sixi=cmmi6
\skewchar\eighti='177 \skewchar\sixi='177

\font\eightsy=cmsy8
\font\sixsy=cmsy6
\skewchar\eightsy='60 \skewchar\sixsy='60

\font\eightbf=cmbx8
\font\sixbf=cmbx6

\font\eighttt=cmtt8

\hyphenchar\tentt=-1 
\hyphenchar\eighttt=-1
 
\font\eightsl=cmsl8

\font\eightit=cmti8

\newskip\ttglue

\def\eightpoint{\def\rm{\fam0\eightrm}%
  \textfont0=\eightrm \scriptfont0=\sixrm \scriptscriptfont0=\fiverm
  \textfont1=\eighti \scriptfont1=\sixi \scriptscriptfont1=\fivei
  \textfont2=\eightsy \scriptfont2=\sixsy \scriptscriptfont2=\fivesy
  \textfont3=\tenex \scriptfont3=\tenex \scriptscriptfont3=\tenex
  \def\it{\fam\itfam\eightit}%
  \textfont\itfam=\eightit
  \def\sl{\fam\slfam\eightsl}%
  \textfont\slfam=\eightsl
  \def\bf{\fam\bffam\eightbf}%
  \textfont\bffam=\eightbf \scriptfont\bffam=\sixbf
   \scriptscriptfont\bffam=\fivebf
  \def\tt{\fam\ttfam\eighttt}%
  \textfont\ttfam=\eighttt
  \tt \ttglue=.5em plus.25em minus.15em
  \normalbaselineskip=9pt
  \let\sc=\sixrm
  \let\big=\eightbig
  \normalbaselines\rm}
%
%
\def\starttitleauthor{}
\def\endtitleauthor{}
\def\author#1{\smallskip\centerline{\allcaps#1}}
\def\institution#1{\centerline{\eightpoint #1}}

%
%
\newcount\sectionnum
\newcount\subsectionnum
\newcount\subsubsectionnum
\newcount\eqnum
\global\sectionnum=0
\global\subsectionnum=0
\global\subsubsectionnum=0
\global\eqnum=0
\def\section#1{\global\advance\sectionnum by 1\global\subsectionnum=0
    \goodbreak\bigskip\centerline{\the\sectionnum. #1}
    \nobreak\medskip}
\def\subsection#1{\global\advance\subsectionnum by 1 \goodbreak\medskip
    \centerline{\the\sectionnum.\the\subsectionnum. \it #1}
    \nobreak\smallskip}
\def\subsubsection#1{\global\advance\subsubsectionnum by 1 \goodbreak\medskip
    \centerline{\the\sectionnum.\the\subsectionnum.\the\subsubsectionnum. \it #1}
    \nobreak\smallskip}
%
%
\def\puteqnum{\global\advance\eqnum by 1 \eqno{(\the\eqnum)}}
%
%
\def\abstract#1{\centerline{\sl ABSTRACT}\smallskip
    {\narrower\eightpoint\baselineskip=11pt\par #1\par}}
%
%
\def\subjectheadings#1{\smallskip{\narrower\eightpoint\par\hangindent 8em
      \noindent\hskip 8em
      \llap{\it Subject headings: }#1\par}\bigskip}
%
%
\footline={\ifnum\pageno>1{\hss\tenrm\folio\hss}
           \else{\ifnum\pageno<-2{\hss\tenrm\folio\hss}
                 \else{\hfil}
                 \fi}
           \fi}
%
%
%
%
%
%
\def\smallup#1{\raise 1.0ex\hbox{\sixrm #1}}
\def\smallvfootnote#1#2{\vfootnote{{\eightrm\raise 1.0ex\hbox{\sixrm#1}}}
       {\baselineskip=10pt\eightrm #2}}
%
%
\def\figpar{\par\noindent\hangindent=1.5em}
\def\figcapt#1{\filbreak\smallskip\figpar{\bf Figure #1}\ --\ }
\def\startfigcapt{\vfill\eject
     \centerline{\bf Figure Captions}\nobreak\medskip}
\def\endfigcapt{\vfill\eject}
%
%
\def\starttables{\vfill\eject}
\def\endtables{\vfill\eject}
%
%

%
%

\def\ltsima{$\; \buildrel < \over \sim \;$}
\def\ltsim{\lower.5ex\hbox{\ltsima}}
\def\gtsima{$\; \buildrel > \over \sim \;$}
\def\gtsim{\lower.5ex\hbox{\gtsima}}

%
%
\def \K         {{\rm\,K}}
\def \Msun      {{\rm\,M}_{\odot}}

\def \kmsmpc    {{\rm\,km\ s^{-1}\ Mpc^{-1}}}

\def \hmpc      {{\rm\,h^{-1}\,Mpc}}

\def \kpc       {{\rm\,kpc}}

\def \cc        {{\rm\,cm^{-3}}}

\def \erg       {{\rm\,erg}}

\def \s         {{\rm\,s}}
%
%
\def\startreferences{\vfill\eject\centerline{\bf References}}
\def\endreferences{\vfill\eject}
%
%
\def\refpar{\par\noindent\hangindent=1.5em\frenchspacing}
\def\bref{\refpar}
\def\jref#1;#2;#3;#4;#5;#6.{\refpar #1 #2, #3 #4, #5}
\def\jrefabbrev#1;#2;#3;#4;#5;#6;#7.{\refpar #1 #2, #3 #4, #5 (#7)}
\def\jrefprivcomm#1;#2;#3;#4;#5;#6.{\refpar #1 #2, private
       communication}
\def\jrefinprep#1;#2;#3;#4;#5;#6.{\refpar #1 #2, in preparation}
\def\jrefpreprint#1;#2;#3;#4;#5;#6.{\refpar #1 #2, preprint}
\def\jrefsub#1;#2;#3;#4;#5;#6.{\refpar #1 #2, #3, submitted}
\def\jrefpress#1;#2;#3;#4;#5;#6.{\refpar #1 #2, #3, in press}
%
%
\def\ana{A\&A}
\def\aj{AJ}
\def\apj{ApJ}
\def\apjl{ApJ}
\def\apjs{ApJS}
\def\araa{ARA\&A}

\def\fcp{Fund. Cosmic Phys.}

\def\mnras{MNRAS}

%
%
\def\xten#1{\times 10^{#1}}
\def\etal{et al{.}\ }
\def\np{\vfill\eject}

\def\ie{i{.}e{.},\ }

\def\eg{e{.}g{.},\ }

\def\numsym{\raise 0.4em\hbox{${\scriptstyle\#}$}}

%
%
%
%
%
\bigskip
\starttitleauthor
\centerline{\sslarge Galaxy Formation in Hierarchical Models}
\bigskip
\author{F J Summers}
\institution{Princeton University Observatory,
   Peyton Hall, Princeton, NJ 08544}
\vskip -0.1truecm
\institution{summers@astro.princeton.edu}
\endtitleauthor
%
%
\bigskip\bigskip

\abstract{
High resolution gravity plus smoothed particle hydrodynamics simulations
are used to study the formation of galaxies within the context of
hierarchical structure formation. The simulations have sufficient
dynamic range to resolve from ten kpc scale galactic disks up to many
Mpc scale filaments. Over this range of scales, we find that
hierarchical structure development proceeds through a series of
increasingly larger filamentary collapses. The well resolved simulated
galaxies contain hundreds to thousands of particles and have varied
morphologies covering the entire expected range from disks to tidally
distorted objects. The epoch of galaxy formation occurs early, about
redshift 2.5 for $10^{12}\Msun$ galaxies. Hierarchical formation
naturally produces correlations among the mass, age, morphology, and
local density of galaxies which match the trends in the observed
morphology--density relation. We also describe a method of spiral
galaxy formation in which galactic disks form through the discrete
accretion of gas clouds which transport most of the angular momentum to
the inner regions. Such a process is characteristic of the somewhat
chaotic nature of hierarchical structure formation where simple
analytical ideas of spherical collapse appear incongruous.
}

%
%
\subjectheadings{cosmology: large-scale structure of universe ---
   galaxies: clustering --- galaxies: formation}
 
%
%
\bigskip\bigskip

\section{Introduction}

In a previous paper (Evrard, Summers, \& Davis 1994, hereafter Paper
I), we demonstrated the ability to resolve galaxies within
cosmological simulations. In particular, we established that an
abundant population of galaxy-like objects are easily defined, that
these simulated galaxies have sizes and masses characteristic of real
galaxies, and that galactic disks can form naturally from hierarchical
initial conditions.  This paper extends that work by examining the
processes of galaxy formation more closely.

The simple analytic picture of galaxy formation considers an individual
collapsing gas cloud (\eg Eggen, Lynden-Bell, \& Sandage 1962). A cloud
is Jeans unstable to collapse when its gravitational collapse time scale
becomes shorter than the timescale for pressure response.  When the
timescale for cooling the gas is shorter than the free fall time scale,
the cloud can fragment into smaller pieces, presumably leading to
eventual star formation (Gott \& Thuan 1976; Rees \& Ostriker 1977; Silk
1977). The timing of star formation during the collapse is important for
the morphology of the galaxy. Spiral galaxies have disks of young stars
with low velocity dispersions and star formation there must occur after
the disk has formed. In ellipticals, with old stellar populations and
shapes determined by large velocity dispersions, star formation must
occur early in the process. It has been suggested for some time that
ellipticals may form via mergers and interactions rather than by
single cloud collapse (Toomre \& Toomre 1972; Barnes \& Hernquist
1992).

Disk formation is seen as the natural result of collapse of a cloud with
non-zero angular momentum. Centrifugal forces of rotation lead to
preferential collapse along the direction parallel to the rotation axis.
The cloud must remain gaseous, hence collisional, during this
collapse for a thin disk to form. The collapse in the plane of the disk
continues until the disk is rotationally supported against the
gravitational potential. The caveat to this picture is
the creation of the initial angular momentum of the cloud and large
scale tidal torques are often invoked to explain this (see \eg Efstathiou
\& Silk 1983).

Hierarchical structure formation will alter this basic picture
considerably. The simplistic idea of a coherent collapse of a single
cloud gives way to a picture in which small scale objects collapse
first, and then merge into successively larger objects. The relevant
Jeans and cooling criteria address the current level of the hierarchy
and not the entire collapsing region. The timescales between each
stage of building this hierarchy are sufficiently close to the
collapse timescales that one does not know whether a level of
structure will reach an equilibrium before it is incorporated into the
next higher level. The emphasis on merging might interfere with disk
formation or disrupt already formed disks (Paper I; Toth \& Ostriker
1992; Navarro, Frenk, \& White 1994). Hierarchical construction also
may mean that large mass ellipticals are assembled too late to fit
with observations.

There have been many analytical and semi-analytical treatments of
hierarchical galaxy formation (\eg White \& Frenk 1991; Kauffmann,
White, \& Guiderdoni 1993; Lacey \& Cole 1993). A statistical
description of galaxy formation can be created which reproduces many
of the observations.  Some limitations of these methods are that the
important processes must be identified in advance and that it is
difficult to incorporate complex feedback processes. Additionally, one
does not obtain a detailed look at the processes or gain insight into
the dynamical aspect.

Numerical simulations have become a useful tool for examining the
interaction of processes in a dynamical setting. Simulations of
isolated galaxy collapses (\eg Katz \& Gunn 1991; Steinmetz \& Mueller
1994) require simplifying assumptions such as spherical clouds and
solid body rotation to create initial conditions. Compared to the
present work, other simulations which start with cosmological initial
conditions have either not had the resolution to consider morphologies
(Katz, Hernquist, \& Weinberg 1992), or have studied a limited number of
targeted objects (Navarro \& White 1994). The above references and Paper
I have shown that gas can cool efficiently (perhaps too efficiently) to
galactic overdensities and that galactic disks can form naturally in
hierarchical models. Such work, including this paper, is limited by an
incomplete physical model and limited dynamic range. The goal of
resolving galaxies within a region large enough to statistically test
cosmological models and including the diverse physical processes
required has fostered several supercomputing research groups.

This paper is organized as follows. \S2 presents a brief overview of
the simulation, examines some characteristics of hierarchical
formation, and demonstrates the dynamic range available in the
numerical model. We then explore the various simulated galaxy
morphologies and their dependence upon environment. In \S4, we
consider the timing of galaxy formation and its relation to other
galaxy characteristics. The detailed description of the formation of a
galaxy with a strong disk and the implications for spiral galaxy
formation are covered in \S5.  Discussion of our results within larger
contexts rounds out the paper.

\section{Simulation Overview}

The numerical details of the simulation were presented in Paper I. Only
a brief description will be given here.  The simulation follows the
coupled evolution of dark matter and baryonic particle fluids with the
P3MSPH code (Summers 1993; Evrard 1988). Initial conditions are chosen
from a CDM power spectrum with cosmological density parameter
$\Omega=1$, present Hubble parameter $h = 0.5$ in units of $100\kmsmpc$,
and normalization of $\sigma_8=0.6$, where $\sigma_8$ is the linearly
predicted rms fluctuation within 8$\hmpc$ radius spheres.  262,144
particles per species are tracked within a cubical region with comoving
side length, $L$, of 16 Mpc.  The simulated region is constrained to
contain a mass concentration characteristic of a poor cluster of
galaxies in the center. The timestep is 6.1 Myr, the resolution scale is
$\sim 20/(1+z)\kpc$, and the particle masses are $9.72\xten{8}\Msun$ and
$1.08\xten{8}$ for the dark matter and baryonic particles, respectively.
The baryonic species is governed by the equations of smoothed particle
hydrodynamics (SPH) which includes radiative cooling for a primordial
composition gas.  The simulation is designed to cover the smaller scales
of cosmology ($\sim 10$ Mpc) down to the larger scales of galaxies
($\sim 10$ kpc).

Specific limitations of the simulation were also described in Paper I
(see \S 2.2), but a few comments bear mention here as well. First,
resolution constraints do not allow us to see structure forming on
scales below $3\xten{10}\Msun$. The first resolvable objects are about
the size of the Large Magellanic Cloud. In as much as the information
from smaller scales is forgotten when the resolvable structures form,
our simulation picks up the formation in progress. Second, the assumed
cosmology does not match the normalization predicted by the COBE
observations (Efstathiou, Bond, \& White 1992).  Accordingly, this paper
does not attempt to produce global tests of the CDM model, but rather
focuses on the characteristic processes in any hierarchical structure
formation model. The important point is that the power spectrum falls
roughly as $P(k)\propto k^{-2}$ on galaxy scales.  The simulation can
essentially be scaled to match the normalization of a desired model by
considering it to occur at a different epoch. Third, we do not include
radiative heating from an ionizing background.  This energy input would
only affect the smaller objects and the very efficient cooling found in
the large objects would not be altered (Katz, Weinberg, \& Hernquist
1996). Fourth, and most important, is the omission of star formation.
Our simulation does not identify when star formation occurs during the
collapse process and is strictly correct only during the gaseous phase
of collapse. Delving into star formation will add a new set of
parameters to the simulation model and will be explored in a subsequent
paper.  It is important to define the situation with well known physics
before pushing into areas that are less well constrained on these scales.

Figure 1 presents the time development of structure in the simulation. A
slice of dimensions $L\times L\times 0.2L$ shows the principle collapse
plane of the poor cluster of galaxies.  The dominant object is the
cluster halo in the central region. It contains a dark matter mass of
about $5.5\xten{13}\Msun$ within a virial radius of $486\kpc$ (defined
by an overdensity of 200) at the end of the simulation. A group
of galaxies forms above and to the right of center. It has a final
virial mass of $1.6\xten{13}\Msun$ within $319\kpc$. No doubt if the
simulation were continued, this group would eventually merge with the
central cluster. Many galaxy scale structures appear throughout the box. 
Most are concentrated in filamentary structures, but some appear in
relatively low density environments (though one must be careful because
these plots are projected slices). The sizes of the filaments and voids are
limited only by the size of the simulation.

The development of linear structures, generically called filaments,
illustrates the hierarchical nature of the process.  At early times
there are many short segments connected in a kinked fashion to a few
longer sections.  As time progresses, the kinked regions collapse into bound
structures and the filaments become longer and straighter. Additionally,
collapse occurs within the filaments themselves and their appearance
changes from continuous to dotted to clumpy. It has been noted before
that groups and clusters form at the vertex of several filaments, but
note here how the formation of the cluster helps strengthen and define
the alignment of the filaments. It is a dual process that creates both
the cluster and the aligned filaments in tandem. The generalization of
this idea is that, at any given stage of structure development,
filaments should connect the dominant collapsed structures.  These
filaments will reflect the progression of hierarchical collapse in their
transition from shorter to longer, kinked to straight, and continuous to
clumpy.

When concentrating on the high density structures, one follows the
collapse of positive mass perturbations and the process of collecting
them together into larger structures.  Alternatively, one can reverse
the foreground and background by following the expansion of negative
mass fluctuations, or voids. The theoretical advantage is that void
regions obey the linear theory equations. The practical advantage can be
seen by tracking the voids in Figure 1. While the voids at the end of the
simulation are apparent back to the initial conditions, the final
positions of the overdense structures do not correlate well with their
initial positions. This suggests that describing large scale structure
by expansion and merging of voids might be analytically fruitful because
the centers of voids are relatively stationary. It also illustrates the
problems entailed in reconstruction methods for determining the initial
perturbation field (Weinberg 1992). Many stages of the hierarchy have
become mixed and are cloaked by the final cluster halo.

Although only the dark matter is shown in Figure 1, the baryons trace
the same structures on these scales (see Paper I, Fig. 3).  Above 1 Mpc,
there is no evidence for segregation of the baryons and the dark
matter.  Indeed, as the collapse is pressureless until a shock is
reached, no segregation should be expected.  The Mpc scale demarks the
point where hydrodynamic effects become of the same order as
gravitational effects for this simulation.  On smaller scales,
hydrodynamics is required for a correct description.

This assertion is punctuated by the dynamic range comparison in Figure
2. The top row exemplifies why dark matter simulations are unsatisfying
for galaxy formation studies.  The dark matter structures that look
interesting in one panel appear bland in the panel to the right. The 400
kpc panel reveals that the cluster halo has almost no sub-structure and
the galactic halo in the 70 kpc panel is difficult to identify. In
contrast, the baryons show a richness of structure on all three levels.
The cluster halo is populated by an assortment of simulated galaxies.
The galactic scale dark matter halo surrounds a collapsed baryonic
galaxy much as envisioned by, for example, White \& Rees (1978).  This
figure affirms the simulation aim of modelling structure from a several
kpc to a several Mpc.

The results of this section show that the simulation is adequate to the task
of addressing galaxy formation in a cosmological context.
The large scales of the simulation exhibit the characteristics of
hierarchical structure formation.  The wide dynamic range and high spatial
resolution allow one to connect to galaxy scales.
We can now look at some characteristics of the simulated galaxies and
at their formation processes.

\section{Morphology of Simulated Galaxies}

Paper I describes the straightforward method of identifying simulated
galaxies as high density baryonic objects using a friends-of-friends
grouping algorithm. We also noted the formation of a substantial
population of disk structures. In this section, we would like to examine
further the morphological characteristics of the galaxy population.
Shapes can be classified for the larger objects and compared to the
observed type of elliptical, spiral, and irregular. With a resolution of
order 10 kpc, only the large scale galaxy structures can be seen and
little information will be provided on internal structure. 
Additionally, the SPH method imposes some strict limitations.

The most limiting aspect is that the baryons in the simulation are
always treated as gaseous. A main effect of thermal and artificial
pressure is to prevent interpenetration of gas clouds by shocking. In
regions of high density and high pressure, not only strongly convergent
flows, but also random motions, will be damped out.  Gas dynamics
strives for smooth flow fields with little dispersion.  The two
solutions are simply translation and rotation. One should expect the
galaxy objects to have a bulk velocity and/or a rotational velocity with
small internal dispersions.

This expectation has several consequences for morphological studies.
Elliptical galaxies, whose shapes derive from the velocity dispersions of
stars, will not be seen. Such objects in this simulation would have their
internal motions damped and collapse into high density lumps at the
resolution scale. The rotational motions of spirals, however, can be
simulated. Angular momentum in collapsing gas clouds will be preserved
and disk structures can form. These objects, too, will have their
dispersions reduced and will be smaller. Any halo component in the
spirals will collapse to the unresolved scales. Irregular galaxies will
follow the same pattern: random motions will damp to the resolution
scale and aligned motions, such as rotation or tidal tails, will remain.
Mergers and interactions between galaxies should strip mass either by
ejection during encounter or via linear tidal features. A gaseous galaxy
cannot be dispersed by pumping energy into internal motions. These
limitations are inherent in the computational method. The inclusion of
star formation will ameliorate many of these considerations and
exacerbate others.

The formation of galaxies with disk structures was reported previously.
An example of a disk galaxy with an infalling satellite is presented in
Figure 3. The large object has a baryonic mass of $1.5\xten{11}\Msun$
with a disk diameter of 18 kpc and the satellite contains
$1.2\xten{10}\Msun$ and is at about 40 kpc away.  The dense central
cores are partly numerical artifacts because the gas particles with
random motions (\ie the ones that might form a stellar bulge) have their
orbits damped to this scale. Also note that the angular momentum and
hydrodynamics have produced disks thinner than the softening length; a
feature which aids recognition, but does not represent the true
resolution scale of the disk.

An example of a tidal encounter is
shown in Figure 4. The projection is face-on to the disk and the
satellite's orbit is inclined about 30 degrees to this plane. As the
satellite begins its orbit, leading and trailing tidal tails develop.
The leading tail appears to accrete both onto the disk and into a small
lump during a close pass. A long trailing structure is evident, its
striking thinness is a numerical effect due to the lack of velocity
dispersion in the satellite. Some comparison can be drawn to the
Magellanic clouds where the Magellanic stream has been suggested to have
been tidally stripped (Lin \& Lynden-Bell 1982; Moore \& Davis 1994). 
However, at large radii the simulated satellite has two tails while no
leading tail for the Magellanic clouds is apparent. Within the disk
galaxy, a ring structure appears to form in response to the satellite's
orbit, but our resolution of such structure is marginal.  The simulation
has resolved not only galaxies, but also the interactions between
galaxies and satellites. For those accustomed to cosmological
simulations only producing large scale plots like Figure 1, the detail
in this figure is remarkable.

There are other details in the simulation that should be noted not for
their scientific merit, but for their complete lack of merit.  The left
panel of Figure 5 shows an extreme close-up of the second largest
object. Note the fine structure resembling spiral arms. Such structure
is completely unresolved and totally bogus. The scale of the core region
in this figure is about 3 kpc, well below the resolution limit. On these
scales, the gravitational force becomes constant and the pressure force
drops to zero. The particles may appear to form structure, but it is in
no way physical. The right hand panel of Figure 5 shows the same view on
a scale four times larger.  $3.7\xten{11}\Msun$ of baryons have
collapsed to an amorphous lump about 20 kpc by 10 kpc by 15 kpc. About
75\% of this mass is concentrated within 2 kpc of the center. This is an
example of an unresolved object. If star formation were included, one might
expect these galaxies to be supported by internal motions and resemble
elliptical or lenticular galaxies.
 
All of the expected morphological types are present. Using a three
dimensional visualization program, shapes have been examined for the 35
simulated galaxies with more than 300 particles. Each is classified as a
disk, an unresolved lump, or a merger, and the local environment, cluster
or field, is noted. The results are given in Table 1. The distribution
changes considerably between field and cluster environments. The
fraction of disks decreases from 86\% in the field to 54\% in cluster
environments. These fractions compare to the observed morphology-density
relationship which finds a spiral fraction up to 90\% in the field and
as low as 10\% in the richest clusters (Dressler 1980; Postman \& Geller
1984). Our poor cluster has perhaps a slightly enhanced spiral fraction
relative to observations. Although the two mergers occurring at this
output are found in the field, the disk abundances argue for more
merging in cluster environments.  One of the mergers clearly shows a
satellite impacting a disk face-on and creating large waves throughout
the structure. Many of the other disks have smaller amplitude warps. 
These occurrences mesh well with theoretical predictions that mergers
will disrupt disks and imply more merging occurs where the disk fraction
is low. One object in the central cluster shows a strong tidal
elongation, having just recently passed by the most massive galaxy.
Tidal features are more prevalent on the smaller mass objects in the
simulation and not on these large ones. A diversity of structure is in
evidence, only some of which has there been space to describe in this
section.

\section{Galaxy Formation History}

Having many outputs from the simulation, it is possible to trace the
simulated galaxies throughout their evolution. To do this, we have
examined the particles in each object found by the grouping algorithm
over 21 outputs. The particles are tracked to find out which objects, if
any, they reside in at the previous and next outputs. If a significant
percentage of particles, taken here as 30\%, are found in a previous or
next object, then the two objects are linked together. In this manner
one builds up a network spanning the outputs and can follow the growth
of an object and the merging of objects by tracing the links. We also
tracked the infrequent instances when an object split into two pieces or
disappeared from the object list. The split objects were actually two
objects that made a close pass and were linked together by the grouping
algorithm during the pass. Most often, they merged soon thereafter. 
Objects that disappeared from the list were small mass ones that
temporarily fell below the minimum particle cutoff.

When an object appears on the list for the first time, it is considered
to be newly formed. The top panel of Figure 6 shows a histogram of the
formation of galaxy-like objects with baryonic masses 
$M_B > 3.2\xten{9}\Msun$. The peak of formation at this mass scale
occurs very early, only about $10^9$ years after the Big Bang, with a
roughly exponential decline in formation rate thereafter. Translated to
redshifts, these small objects reach their peak formation rate at
redshift $z=4$. 
Gaussian statistics predicts a tail of the collapse of very small
overdensity perturbations continuing to $z=0$, and such a tail is
consistent with our data.  The objects plotted in the histogram are the
smallest resolvable objects in our simulation. Larger mass scales
collapse later, as the hierarchy builds. Objects with
$M_B > 1.1\xten{11}\Msun$, approximately the baryonic mass of an $L_*$
galaxy for the assumed baryon fraction, have a later peak
formation redshift of $z=2.5$.  This epoch is still only $2\xten{9}$
years after the Big Bang and over $10^{10}$ years before the present day
in this cosmology. Galactic structures can form in a relatively short
time compared to their ages.


The bottom panel of Figure 6 shows the merging histogram for these
objects. A merger is counted when two objects trace to the same object
at the next output. If more than two objects merge into one object,
extra events are counted. As expected for hierarchical clustering, the
peak of merging is delayed with respect to the peak of formation. Notice
that the merging peak occurs at about the same time as the formation
peak of objects 30 times more massive. If merging were the dominant
process for increasing mass, one would expect the merging peak to
coincide with the formation peak of objects only a few times more
massive. Clearly, accretion of mass which is not in collapsed objects is
more important for building these objects than merging. Merging also
occurs more stochastically than formation with more variance in the
histogram. The peak in the merger rate at $z=2.2$ can be identified with
the collapse of filaments around the central cluster and the formation
of the central object of the smaller group (see Figure 1). The later
merging peaks at $z=1.5$ and $z=1.1$ are correlated with the periods of
heaviest infall through the center of the central cluster. These
observations suggest that merging is heavily linked to group and cluster
formation, while accretion is the dominant process for adding mass to
galaxies.

The above discussion concerns the initial formation of objects at one
mass scale and does not relate the objects to their final states. 
Certainly, one does not expect a $10^{10}\Msun$ object at $z=5$ to have
no mass accretion and retain that same mass today. All of the simulated
galaxies at the final output are traced back to their earliest forming
resolved progenitor. Figure 7 shows the resulting correlation between
redshift of formation (or first appearance), $z_f$, and the final mass. 
Such a trend would follow from a simple accretion argument: objects
which form earlier have a longer time to accrete and end up with
larger masses. However, the mass accretion rates for these objects are
not constant over time (Paper I, Fig{.} 13) and such reasoning is
oversimplified. The correlation also reflects the tendency for early
forming objects to be in regions of higher density on larger scales.
Thus, high redshift objects have both more time and more mass with which
to build to larger mass scales.

This relationship between earlier formation and higher mass provides
consistency amongst other results.  Cen \& Ostriker (1993), using a lower
resolution simulation with heuristic galaxy formation, also reported a
trend that the earliest forming objects ended up in the densest regions.
Previously, we reported that the more massive objects are likely to be
found in the denser environments of the simulation (Paper I, Fig{.} 10).
Combining these results and the one above, we see that our high
resolution run confirms the Cen \& Ostriker result and goes further to
show a correlation of both mass and formation redshift with local galaxy
density. Extending these ideas, one notes that larger mass and older age
(higher $z_f$) correlate to earlier Hubble type.  The formation of early
type galaxies predominantly in higher density environments was confirmed
by the morphological studies in \S3. Thus, hierarchical structure
formation naturally agrees with the trends in mass, age, and
morphological type in the observed morphology-density relation.

\section{Disk Formation and Hierarchical Collapse}

Galaxy formation is inherently a complex, dynamical, three dimensional
process. Much of the intuition for the process is gleaned by examining
visualizations of the simulation on a computer screen. A characteristic
pattern of disk formation emerges from studying the process for the well
resolved objects. Disk formation will be presented by studying one
example in detail. This example is typical in that it highlights the
important points in a relatively straightforward way. Much more
information is contained in the three dimensional and kinetic aspects,
but these figures provide the essential flavor of the disk formation
process. Other objects show variations on this theme.

\subsection{The Formation Process}

Figure 8(a,b) provides a time sequence of the collapse of the sixth most
massive galaxy. Within a radius of 50 kpc, the total mass is
$1.2\xten{12}\Msun$ and the baryonic mass is $2.0\xten{11}\Msun$. Note
that the virial radius is much larger at 171 kpc, and contains total
mass $2.4\xten{12}\Msun$ and baryonic mass $2.7\xten{11}\Msun$.  The
panels show only the baryonic particles in 200 kpc cubic regions at
approximately 215 Myr intervals. Velocity vectors are plotted as tails. 
The late evolution of this object has been presented in the tidal
encounter of Figure 4.

Being a large object, a perturbation is apparent at high redshift.  By
$z=6.62$ a strong filament has formed with an underdensity to the left
in this projection (XZ). This collapse region is the middle section of
the long filament in the lower right of the first panel of Figure 1 (a
YZ projection).  Following the evolution in Figure 1, one can see that
the section of filament below this region forms another galaxy and the
section above it flows into the central cluster. The bifurcations of the
velocity field are apparent in the $z=9.0$ panel, but quickly expand
beyond the box size.

Skipping for a moment to Figure 9, we show two projections of a 400 kpc
box about the object at $z=6.62$ with circles around the particles that
end up in the densest region at the final output. The filament segment
is clearly marked in the XZ projection (left panel) and seen end-on in
the XY projection (right panel). Note also that the distribution is
simply connected, but highly asymmetrical, indicating that spherical
clouds are a poor approximation in hierarchical formation.  Indeed, the
XY projection indicates a grid of smaller filamentary structures that
will collapse first and then combine with the main perturbation.

Continuing with Figure 8, the $z=5.28$ and $z=4.41$ panels show the
collapse of the filament segment along its axis.  Such a collapse is
mostly one dimensional and contains little angular momentum.  This
sequence of filament formation, segmentation of the filament, and
collapse of the individual segments is apparent not only on galaxy
scales, but also on much more massive scales in larger volume
cosmological simulations. Hierarchical collapse may be loosely
characterized as a fractal collapse of filaments progressing from small
scales to large.

On the heels of collapse along the filament axis comes the infall of the
smaller nearby structure from the right in the $z=4.41$ through $z=3.32$
panels. This material is from the smaller filaments apparent in Figure
9. Much of the structure has collapsed into several small clumps, of
which the two largest appear at the bottom of the plots. The smaller
clump enters the region from the back right corner while the larger's
orbit is roughly in the plane of the projection. The central object has
little vorticity until these gas clouds fall into orbit, dissipate
inward, and dominate the flow field. By the time the smaller one has
accreted, $z=2.20$, a clear disk with a smooth flow field has been
established. While the large cloud does not impact the inner regions,
its gravitational presence does aid disk formation and the plane of the
disk is at an angle intermediate to the orbital planes of the small and
large clumps. Though not clearly seen in this projection, continuous
infall comes from many different directions, is slowed by shock heating,
and softly accretes onto the disk.

The process shown in this example is common to disk formation in the
simulation. The robust elements are the formation of a filament, the
collapse of the filament, infall of a few dense clouds to build a basic
disk structure, and slow accretion of low density material to strengthen
the disk. The angular momentum arises naturally during the collapse
from the tidal fields of neighboring perturbations. Bifurcations in the
gravity field leave irregularly shaped regions that collapse with
significant angular momentum. Galaxies that do not form disks deviate from
this process. Some form near the vertex of several small filaments and have
mostly radial (small impact parameter) cloud accretion. Others
encounter the hot gas in group or cluster environments and have their
surrounding gas heated to the inefficient cooling regime and/or stripped.

\subsection{The Resulting Structure}

To illustrate the structure of the resulting disk galaxy, Figure 10
shows averaged radial profiles of several quantities out to 100 kpc. The
gas density falls as a power law with $\rho_{\rm gas}\propto r^{-2.8}$
while the total density (not shown) follows $\rho_{\rm tot}\propto
r^{-2.4}$.  The density spike at 60 kpc indicates the satellite galaxy
and is very prominent because the points are mass weighted averages. The
temperature is near the minimum cutoff value in the dense inner regions
and quickly rises to the virial temperature.  The mass profiles show
that the baryons dominate the mass on scales close to the resolution
limits and only reach the cosmic proportion at about 100 kpc.  Last, the
radial dependence of the smoothing length is provided to remind the
reader of the variable resolution of the hydrodynamics. Resolution is
proportional to $\rho^{-{1\over3}}$, making the effective radii of
particles larger in the low density regions. These profiles are very
similar in all the well resolved disk objects and agree with those of
Navarro \& White (1994).

The components of this simulated galaxy form a neat and stable picture.
At the interior is a rotationally supported disk with baryons dominating
the central mass and falling off more sharply than the total mass.
Enveloping this is a $10^6$ K isothermal gas halo which extends ten
times larger than the cooled region and contains about half of the
baryonic mass within the virial radius. Finally, there is a dark matter
halo providing the dominant mass and which is in cosmic proportion with
the baryons at large radii of the system.  The structure of this object
is basically unchanged for another 1800 Myr through the end of the
simulation and presumably would remain so until a merger with a galaxy
of comparable mass or interaction with an intracluster medium could
perturb it.

Such structure is oversimplified by the numerical model. The stellar
bulge is missing for reasons described above and other physics of
galaxies is either smoothed or not included. The size of the disk is
likely decreased from its natural size because of numerical effects,
however, the factor of ten scale difference between the disk and halo
fits theoretical expectations. And the implication that a significant
amount of baryonic mass resides in a gaseous halo is certainly
intriguing. Overall, the correspondence to the analytic picture of a
dark matter dominated galaxy is relatively strong.

\subsection{Analysis of the Process}

The above picture of disk formation is quite different from the simple
spherical collapse model. The distribution of matter is decidedly not
spherical nor uniform. The collapse is also not coherent nor
synchronized, but occurs in many places with simultaneous formation on
several scales.  In addition, only a small homomorphic collapse of the
disk is required for it to be rotationally supported, not the large
factor envisioned by the simple model. Finally, the angular momentum is
not smoothly distributed throughout the collapsing protogalaxy, but, as
shown below, becomes dominated by the motions of a few objects. The
spherical collapse model greatly oversimplifies the rich complexity of
the collapse process.

The development of the angular momentum in the system is interesting.
The origin of angular momentum in galaxies is usually attributed to
tidal torques in the surrounding gravitational potential. One can see
the process occurring early on in the figures. The fracturing of the
gravity field leaves an irregularly shaped and unevenly structured
collapse region that will naturally develop angular momentum when
collapsing in a tidal field. In Figure 11, we show the development of
the spin parameter in the system, defined as $J|E|^{1/2}G^{-1}M^{-5/2}$
where $J$ is the angular momentum, $E$ is the total energy, $M$ is the
mass. Most of the angular momentum in the baryons (top panel) is
concentrated in the two clouds, which show as distinct peaks of the
curves.  When they accrete, their rotation is deposited over the inner
regions. The dark matter spin parameter (bottom panel) remains near the
same magnitude throughout. A small effect from the clouds is apparent,
but it washes away and the entire dark matter curve at $z=1$ is
$\ltsim 0.1$. Figure 11 affirms the visual intuition one gets for the
development of the angular momentum of the system from Figure 8.

The emergent picture of disk formation is one dominated by discrete
accretion of gas clouds. Disks arise from infall of individual dense gas
clouds that carry most of the angular momentum of the system. These
clouds can dominate the flow because they have formed at large radii and
infall with large impact parameters; thereby combining significant mass,
linear momentum, and angular momentum.  Additionally, the matter in the
small filamentary structures begins to collapse together while infalling.
Hence, infall directions, and thus, angular momentum vectors,
are somewhat correlated. Though more mass is accreted in low density
material, this material does not have large, compact momentum and is
slowed before it can directly impact the inner regions. This soft
accretion adds to the mass of the disk, but does not establish a flow
pattern. Galaxy formation in hierarchical models invokes the idea of 
disk formation by discrete accretion.

The emphasis on merging for building a galaxy in hierarchical structure
formation has called into question the ability to form disk structures
and their stability after formation. If the
merger events were numerous, the chance for correlated angular momentum
would be low and late mergers would heat and destroy a disk.  However,
hierarchical structure formation requires only a few significant events
to build the galaxy. The next level would be a merger with an object of
similar size, not a continuing rain of smaller objects. For those
objects in the simulation which do experience many mergers, resolved
disks do not form. We also see encounters where satellite galaxies
impact on already formed disks and strongly disrupt them. The high
fraction of disk morphologies indicates that these are the exceptions,
not the rule. This simulation demonstrates that galactic disks not only
form in abundance within hierarchical models, but also can survive for
several times $10^9$ years.

To engage in a bit of speculation, one may directly identify stages of
the collapse process with the structure in spiral galaxies. The collapse
along the filament axis produces little angular momentum and would
characteristically form a halo component.  Subsequent stellar accretion, which
cannot be differentiated in this simulation, would build a halo in the pattern
of that described by Searle and Zinn (1978).  The cloud infall to establish
a disk would have significant dispersion, some internal to the clouds
and some created in the process of defining a rotation plane, that might
lead naturally to a thick disk. Later gas accretion and settling would be
softer and more coherent: producing a thin disk component. Although we
have neither the physics to simulate a halo nor the resolution to
differentiate disk components, the picture fits with observed galactic
structure and flows naturally from the dynamical processes.

A healthy skepticism will naturally ask how this formation process is
affected by the simulation limitations. The main point on this matter is
that as long as the baryons remain gaseous, the simulation is correct.
The infalling gas clouds have enough linear and angular momentum that
their ram pressure on the gas in the inner regions produces the vorticity
seen. Artificial pressure effects will, if anything, understate the
ram pressure since the clouds will be slowed in advance of shocks by the
finite resolution of the code. For the example above, the densest
baryons will likely have formed stars and thus become collisionless, but
this will only occur in the inner 10 kpc of the forming galaxy and the
central regions of the large cloud. Densities in these regions exceed
$1\cc$ while the small cloud is of order $10^{-2}\cc$.  Star formation
might allow the large cloud to orbit with less drag, but should not
change the path and angular momentum transfer of the small cloud.

If star formation proceeds efficiently in the infalling clouds, it is
difficult to envision how hierarchical structure formation will produce
an abundant population of disks. The linear collapse of filaments does
not produce much angular momentum. The gravitational influence of
satellite orbits would be inefficient at transferring angular momentum,
especially to the inner regions. Angular momentum would arise mostly
from the slow accretion of low density gas.  However, because the
angular momentum of this gas is not well correlated, it may be unable
to dominate the dynamics. It seems imperative that infalling clouds be
gaseous for hierarchical collapse to produce disks.

\section{Discussion}

This paper has presented a coherent picture of galaxy formation within
hierarchical structure formation models.  An early formation epoch is
observed, with the peak formation rate on the $10^{12}\Msun$ scale
occurring at $z\approx 2.5$. Formation rate after peak drops with an
exponential tail while merger rates increase, though mergers tend to be
dominated by cluster interactions. The range of morphologies observable in
the simulation are a good match to observed types and include
rotationally supported disks and extended tidal tails. The structure of the
disk galaxies consists of a gas disk extending tens of kpc surrounded by
both hot gas and dark matter halos out to ten times larger. Hierarchical
structure formation is quite successful at forming reasonable galaxies.

The observed morphology-density relationship appears as a natural result
of hierarchical formation. In Paper I, we noted that massive galaxies
resided in the denser environments of the simulation. We show here a strong
correlation between mass and early collapse, indicating that 
massive objects are older. Finally, direct examination of morphologies shows
that disk galaxies are more prevalent in the field. Equating disks to spirals
and higher mass and older age to ellipticals, the trends in mass, age, and
morphology agree with the observed morphology-density relationship.

A new method for making spiral galaxies arises in hierarchical structure
formation.  The importance of merging was noted by Navarro \& White
(1994) and here we hone the details of the process.  The early stages
are characterized by large and small filamentary structures which
collapse separately. The angular momentum transport is dominated by a
few gas clouds which can punch through a gas halo, directly impact the
inner regions, and establish a flow pattern.  Disruption of the disk
from too many mergers is avoided because only a couple significant
accretion events are required to build a level of the hierarchy. Because
ram pressure is important for transferring angular momentum, the timing
of star formation during this process will be a key ingredient. Further
studies will be needed to probe deeper into this process of disk
formation by discrete accretion.

One strong conclusion is that, in hierarchical structure formation,
there is no such thing as a simple spherical collapse.  The collapse of
a galaxy sized mass is more chaotic than uniform, with separate pieces
undergoing local collapse simultaneously within the global picture. 
Ideas of sphericity and uniform distribution of angular momentum are
inconsistent with the findings.  While the simple model is no doubt
analytically useful, its practical utility when applied to hierarchical
structure formation should be evaluated.

If we indeed live in a hierarchical universe, the above results have
several implications for galaxies. The halos extend approximately 10
times larger than the luminous parts of the disks.  Rotation curves
should be flat or only mildly falling, and certainly not Keplerian,
within radii of about 150 kpc.  Such a halo size estimate fits well with
that predicted from observed mass to light ratios (Bahcall, Lubin, \&
Dorman 1995), although the conclusions of those authors suggest a low
density universe and our simulation assumes a flat universe.  On the
lower bound, HI observations indicate a cool gas extent several times
the optical disk (\eg Cayatte \etal 1994) and the need for a dark matter
halo already several times more massive than the maximal disk at radii
around 30 kpc (\eg Begeman 1989).  The gas in these large halos is
roughly isothermal at a temperature of a few times
$10^6\K$ and contains one third to one half of the baryonic mass within
the virial radius. At densities near $10^{-3}\cc$ the total free-free
emission is about $5\xten{41}\erg\s^{-1}$ peaked at a couple hundred eV.
Such emission is not strong enough to be seen in the bandpasses of
current X-ray satellites (see also Navarro \& White 1994).  In the
simulated galaxies, the halo gas cools in the inner regions and provides
a continual source of mass accretion onto the disk.  However, as the
simulations do not include interactions between the disk and halo
(galactic fountains, ionizing radiation), such accretion is not a firm
prediction. 

The correlation between an object's merging history and its final
morphology is not completely straightforward. Objects with lots of
merging and recent merging do not form disks. A quiescent merger history
for a field object is a good indicator of disk formation. However, in
the cluster, a quiet history does not always lead to a disk. Tidal
perturbations can disrupt disks that have already formed, while heating
and stripping of halo gas via interaction with the intra-cluster medium 
can interrupt accretion onto disks that are forming. 

Probing the morphological information, one can explore the idea that
ellipticals form from mergers of spirals. For the central galaxy of the
cluster, which has the characteristics of a cD, formation is dominated
by a series of merger events that occur early. The objects which merge
do not have time to develop disks. A couple other objects that are
candidates for ellipticals also form from several relatively quick
mergers where disks do not form first. We also see a couple instances
where formed disks are disrupted, though these may be fairly considered
as interacting galaxies.  Thus, the general trend is that our results do
not support the idea that all galaxies form disks with ellipticals
forming as mergers of disks. Hierarchical collapse in dense regions
progresses through the mass scales too quickly and with too much merging
for disk formation. These conclusions are, however, drawn from only a
handful of cases. Further, we have no handle on addressing dwarf
ellipticals.

Having identified some characteristics of hierarchical formation, it is
useful to consider whether such ideas could translate into observables.
Figure 1 suggests that voids are more spatially stable than
overdensities as features of the density field. The method for
reconstructing the initial density field of Weinberg (1992) might
provide additional insight if focused on the low density regions as
well. Section 5 points out that filaments exist on all scales and that
collapse progresses from small to large. One may find a useful
measurement of the level of collapse in either the smallest filament
which has yet to collapse or in the largest filament that has had time
to form. Such a measure would be spatially variable and could provide an
assessment of the local dynamical state in the mildly non-linear regime.
The difficulty would be in extracting real space filaments from redshift
space data.

This study of galaxy formation has provided numerous insights, but leaves many
topics left to pursue. The presence of the cluster in the constrained initial
conditions skews the formation process a bit. The average redshift of
formation is assuredly higher in this overdense region. Perhaps the strong
filaments which intersect at the cluster overemphasize the role of filaments
in the galaxy formation process. Also, we can resolve formation for galaxies
of mass scale $10^{12}\Msun$, the process may differ for smaller masses. Both
of these arguments are countered by noting that filaments are found on all
resolvable scales and appear to be a ubiquitous part of the process.
Resolution and unconstrained initial conditions will help settle these
questions.

Another point is that the simulation finishes at redshift one. The early
formation epoch for galaxies begs the question of what will happen at
late times. Will the galactic disks remain stable for another
$8\xten{9}$ yr? In the simulated region, most of the objects would end
up in the cluster, where disk stability is questionable. For more
general regions, the formation epoch would be later and the process may
proceed slower, but disks would still need to be stable for a
considerable fraction of the Hubble time. As seen in this simulation,
the early fracturing of the gravitational field breaks the mass into
distinct regions which collapse relatively quickly. Small objects near
large galaxies become satellites as seen in the figures.  There does not
appear to be a large reservoir of disrupters available for late time
infall. Small objects which do form later would be in other regions.
These extrapolations of the current simulation favor disk stability with
the acknowledged exception of galaxy-galaxy interactions. A random
region can be evolved to the present day to check the arguments.

The effects of varying the cosmology should not be great.  By focusing
on small scales and not including the global aspects of the cosmological
model, only hierarchical structure formation has been tested here.
Changing the $\sigma_8$ normalization will essentially just rescale the
redshifts assigned to the simulation and perhaps speed up or slow down
the processes. Using a low density universe may have the largest effect
in that it forces structure formation to occur early and then suppresses
further growth. Such suppression may be required to maintain stable
disks to the present day if the arguments of the preceding paragraph do
not hold or if galaxy mergers are too frequent. Also, the reduced total
mass and larger baryon fraction in low density and cosmological constant
dominated universes will favor smaller galaxy halos and a quieter
velocity field for the same luminosity. These changes may alter such
aspects as the relative importance of interacting galaxies, but the
basic hierarchical process should remain.

The ability to address simulations of galaxy formation within a
cosmological context has been attained only recently. This study
presents only a few of the qualitative and quantitative insights into
the galaxy formation process that can be gained. The prospects for
examining a wide range of characteristics of the galaxy population as it
develops over a Hubble time are quite good.

\bigskip

The author wishes to thank Gus Evrard, Marc Davis, and David Spergel for
helpful discussions. Support for this work was provided by NSF grant
AST-8915633 and NASA grants NAGW-2448 and NAG5 2759. Computing resources
provided by the San Diego Supercomputing Center and the Pittsburgh
Supercomputing Center are gratefully acknowledged.

%
%

\startreferences

\jref 
Bahcall, N. A., Lubin, L., \& Dorman, V.;1995;\apj;447;L81;L85. 

\jref
Barnes, J. E., \& Hernquist, L.;1992;\araa;30;705;742.

\jref
Begeman, K.;1989;\ana;223;47;60.

\jref
Cayatte, V., Kotanyi, C., Balkowski, C., \& van Gorkom, J.
H.;1994;\aj;107;1003;1017.

\jref 
Cen, R., \& Ostriker, J. P.;1993;\apj;417;415;426.

\jref
Dressler, A.;1980;\apj;236;351;.

\jref
Efstathiou, G., Bond, J. R., \& White, S. D. M.;1992;\mnras;258;1p;6p.

\jref
Efstathiou, G., \& Silk, J.;1983;\fcp;9;1;138.

\jref
Eggen, O. J., Lynden-Bell, D., \& Sandage, A. R.;1962;\apj;136;748;766.

\jref
Evrard, A. E.;1988;\mnras;235;911;.

\jrefabbrev
Evrard, A. E., Summers, F J, \& Davis, M.;1994;\apj;422;11;36;Paper I.

\jref
Gott, J. R., \& Thuan, T. X.;1976;\apj;204;649;.

\jref
Katz, N., Hernquist, L., \& Weinberg, D. H.;1992;\apjl;399;L109;L112.

\jref
Katz, N., \& Gunn, J. E.;1991;\apj;377;365;381.

\jrefsub
Katz, N., Weinberg, D. H., \& Hernquist, L.;1996;\apjs;;;.

\jref
Kauffmann, G., White, S. D. M., \& Guiderdoni, B.;1993;\mnras;264;201;.

\jref
Lacey, C., \& Cole, S.;1993;\mnras;262;627;. 

\jref
Lin, D. N. C., \& Lynden-Bell, D.;1982;\mnras;198;707;.

\jref 
Moore, B., \& Davis, M.;1994;\mnras;270;209;221.

\jref
Navarro, J. F., Frenk, C. S., \& White, S. D. M.;1994;\mnras;267;L1;L3.

\jref
Navarro, J. F., \& White, S. D. M.;1994;\mnras;267;401;412.

\jref
Postman, M., \& Geller, M. J.;1984;\apj;281;95;99.

\jref
Rees, M. J., \& Ostriker, J. P.;1977;\apj;179;541;.

\jref 
Searle, L., \& Zinn, R.;1978;\apj;225;357;379.

\jref
Silk, J. I.;1977;\apj;211;638;.

\jref
Steinmetz, M., \& Mueller, E.;1994;\ana;281;L97;L100. 

\bref
Summers, F J 1993, Ph{.}D. Thesis, University of California at
Berkeley.

\jref
Toomre, A., \& Toomre, J.;\apj;1972;178;623;.

\jref
T\'oth, G., \& Ostriker, J. P.;1992;\apj;389;5;26.

\jref
Weinberg, D. H.;1992;\mnras;254;315;342.

\jref
White, S. D. M. \& Frenk, C. S.;1991;\apj;379;52;79.

\jref
White, S. D. M. \& Rees, M.;1978;\mnras;183;341;358.

\endreferences

\starttables

\centerline{\bf Table 1 }
\smallskip
\centerline { Morphological Fractions by Environment }
\bigskip
\vbox{\hbox to \hsize{\hfil\vbox{\halign
  {#\hfil&\quad\hfil#\hfil&\quad\hfil#\hfil\cr
\noalign{\hrule}\cr
\noalign{\smallskip}
\noalign{\hrule}\cr
\noalign{\medskip}
 & Field & Group/Cluster \cr
Morphology & \# \quad\hfil \% & \# \hfil \% \cr
\noalign{\smallskip}
\noalign{\hrule}\cr
\noalign{\medskip}
Disk & 19 \quad\hfil 86 & 7 \hfil 54 \cr
Unresolved & 1 \quad\hfil 5 & 6 \hfil 46 \cr
Merger & 2 \quad\hfil 9 & 0 \hfil 0 \cr
\noalign{\medskip}
\noalign{\hrule}\cr
\noalign{\medskip}}}\hfil}}

\endtables

%
%

\startfigcapt

\figcapt{1} Time development of structure in the simulation. Positions
of the dark matter particles within a slice of dimensions $L\times
L\times 0.2L$ are shown at six output times. The slice contains the
principle collapse plane of the cluster that forms in the central
region. For clarity, only one fourth of the particles are shown.

\figcapt{2} Dynamic range of the simulation. Plots of the dark matter
(top row) and the baryons (bottom row) are shown in three successively
enlarged regions at the end of the simulation. The left panel has
dimensions $0.4L\times 0.4L\times 0.1L$ and shows only one fourth of the
particles.  The center and right panels are cubical with side lengths of
0.057L, and 0.01L, show all particles, and are enlargements of the boxed
area in the panel to their left. Physical dimensions of the region
widths at $z=1$ are given on the plot.  The left panel is a projection
onto the YZ plane while the center and right panels are projections onto
the XZ plane.

\figcapt{3}
A disk galaxy with an infalling satellite in the simulation. All
baryonic particles in the region are shown and perspective has been
added by a visualization program. The disk is approximately 40 kpc
across.

\figcapt{4}
Tidal encounter sequence of satellite infall. The sequence begins in the
upper left and proceeds left to right in each row. All baryonic
particles in the region are plotted in a projection face-on to the disk.
Panels are not equally spaced in time, but cover a total span of about
850 million years.

\figcapt{5}
Unresolved hydrodynamic effects. The left hand panel shows the baryonic
particles at the center of the second most massive galaxy in a cubical
region 9 kpc on a side. The structure in this picture occurs below the
resolution scale of the simulation and must be ignored. The right hand
panel shows the same projection in a cubical region 40 kpc on a side.

\figcapt{6}
Galaxy formation and merging histograms. The top panel shows the number
of new galaxies found by the grouping algorithm at each output. The
bottom panel charts the number of merger events (see text).

\figcapt{7}
Correlation between formation redshift and mass. The redshift when a
galaxy first appears in the simulation is plotted versus its final
baryonic mass for the 215 galaxies found at the final output.

\figcapt{8}
Time sequence of formation of the sixth most massive galaxy.
Each panel shows the baryonic particles in a 200 kpc cubic region for
the redshift indicated. The time interval between panels is about 215
Myr. Velocities are plotted as tails and are scaled such that
1000 km/s equals 0.15 of the box length or 30 kpc. 

\figcapt{9}
This figure shows a larger view of the $z=6.62$ panel of Figure 8(a).
All baryonic particles within a 400 kpc cubic region are plotted and the
particles which reside in the galaxy \numsym 6 at $z=1.01$ are marked by
larger points. The left panel shows the same projection as Figure 8 (XZ)
and the right panel shows a view approximately down the filament axis
(XY).

\figcapt{10}
Radial profiles of several quantities for galaxy \numsym 6 at $z=1.83$.
Values for density, temperature, and smoothing length are mass weighted
average values in each bin. Mass profiles are mass interior to the bin
and are marked by crosses for the baryons and circles for the dark
matter.

\figcapt{11} Spin parameter development. The dimensionless spin
parameter (see text for definition) is plotted as a function of radius at
several redshifts. The top panel is calculated from the baryons, while
the bottom panel displays the curves for the dark matter.

\endfigcapt

%

%
%

\null
\vskip 1.0truein
\centerline{\bf Figures -- Electronic Preprint Version}
\bigskip

For the electronic preprint, the large figure files have been replaced
with 100 dpi resolution scanned versions.
Such processing reduces the file size and the printing time required while
also increasing the compressibility of the files for electronic transfer.
In this paper, the following
figures have been replaced by scanned versions: 1, 2, 3, 4, 5, 8a, 8b, 9.
Full resolution figures can be obtained
from the author or from the WWW site:
``http://astro.princeton.edu/\~{ }summers''.

\np

\psfig{file=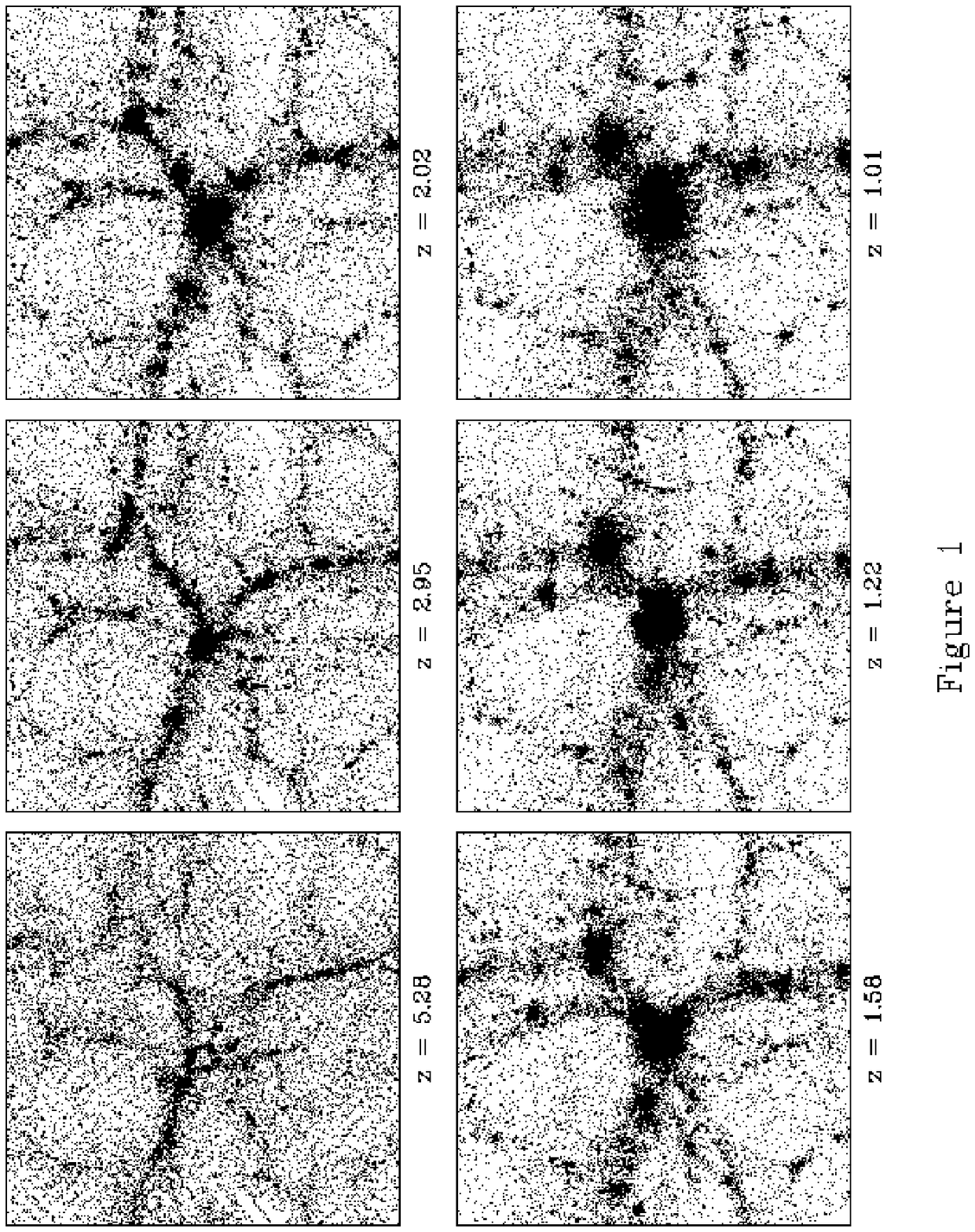,width=560truebp}

\np

\psfig{file=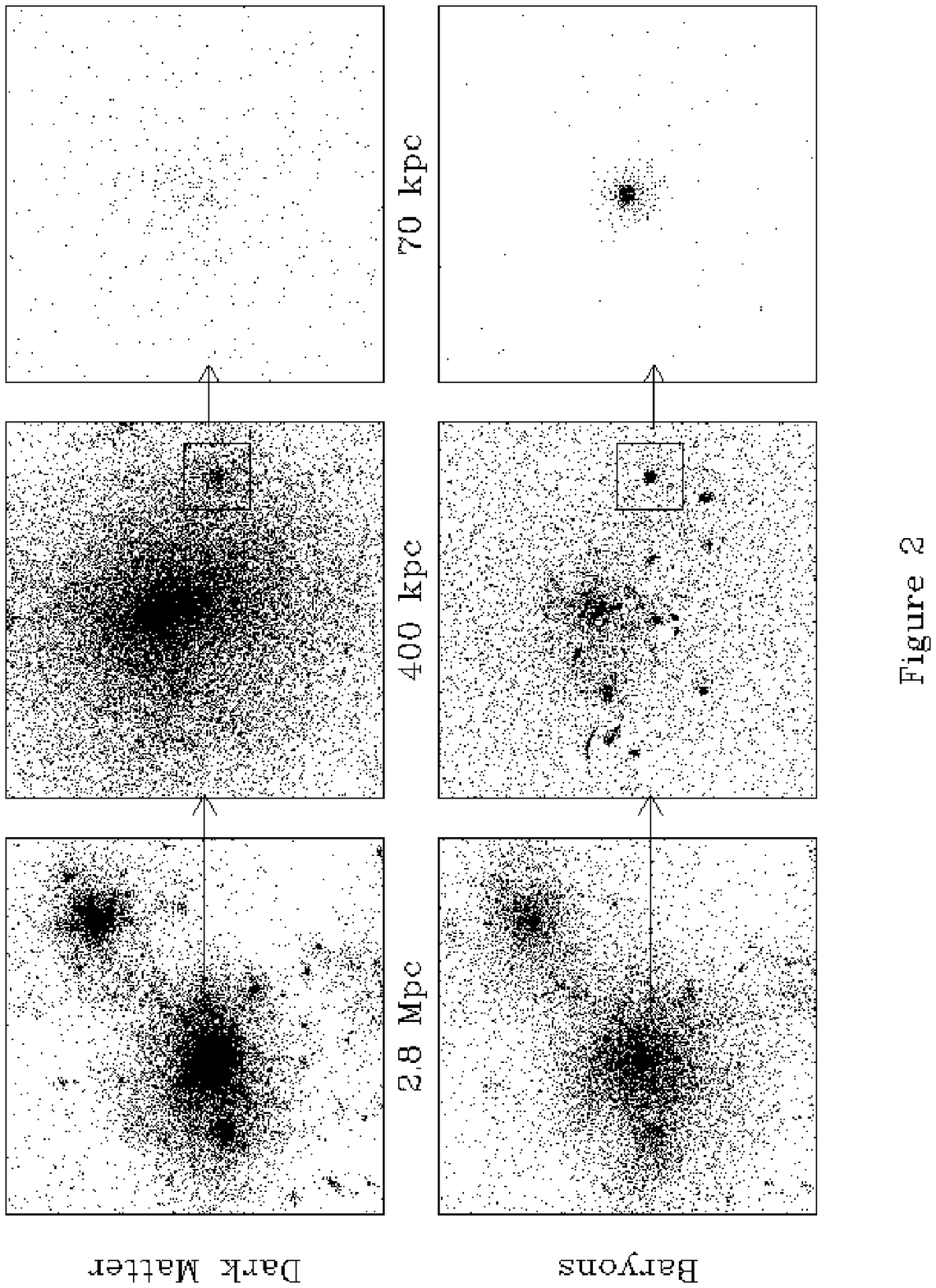,width=560truebp}

\np

\psfig{file=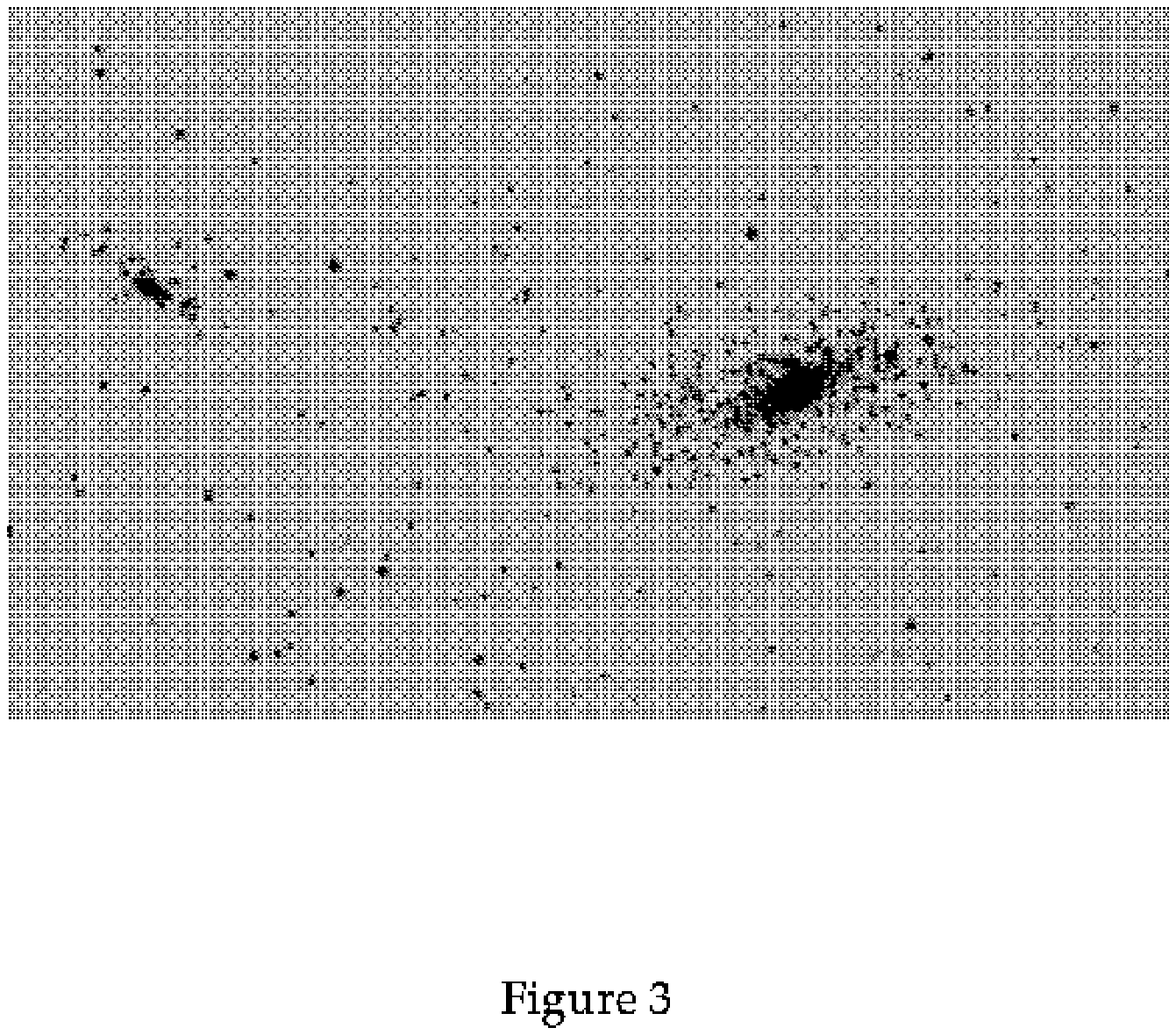,width=560truebp}

\np

\psfig{file=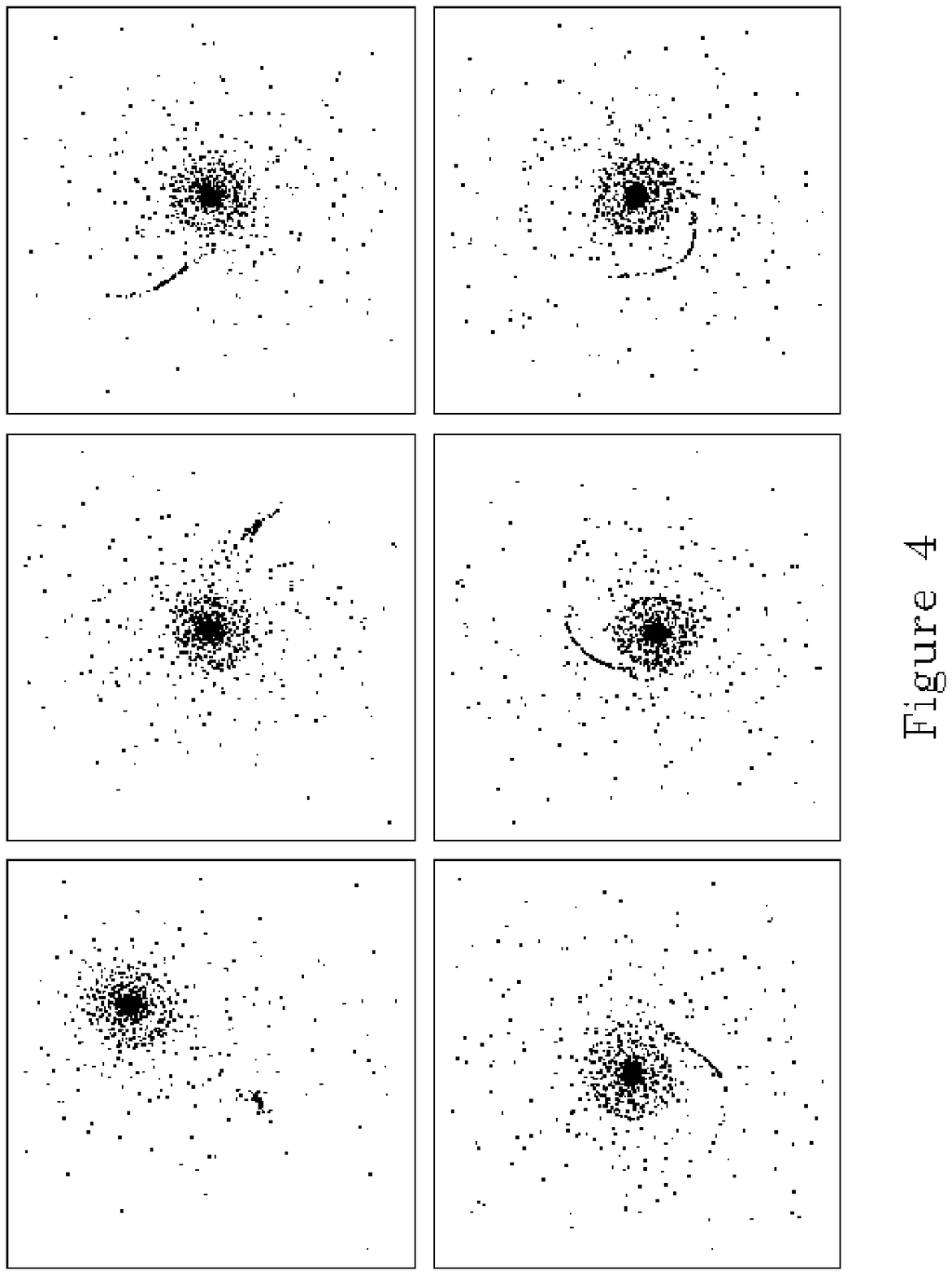,width=560truebp}

\np

\psfig{file=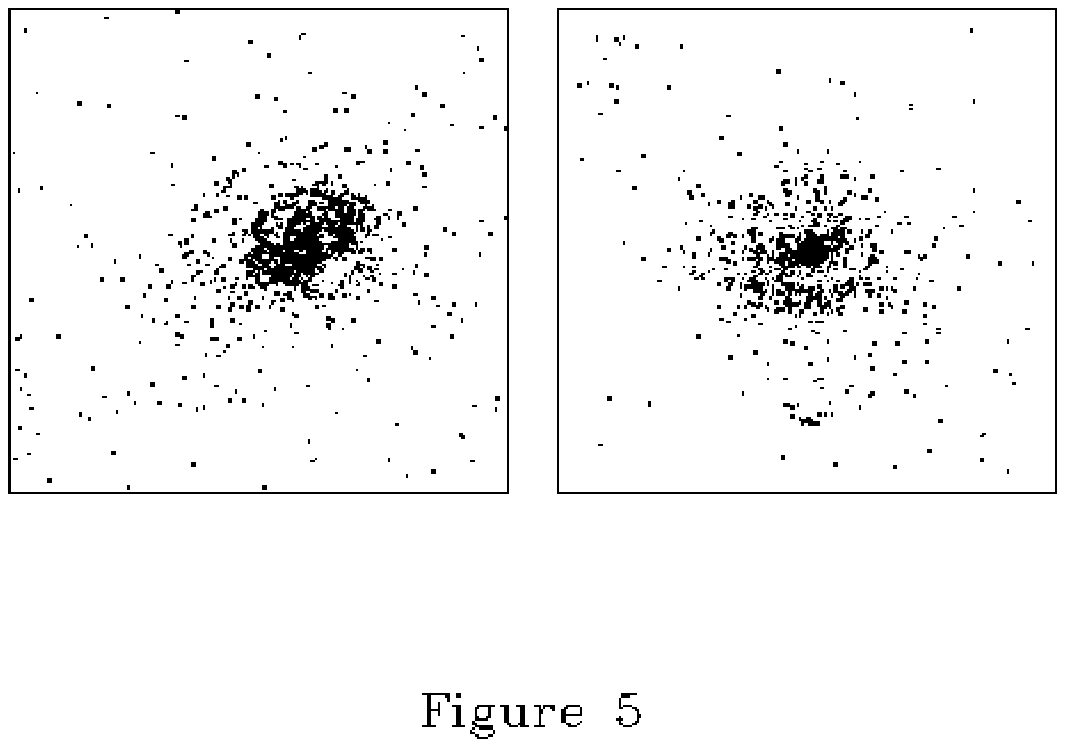,width=560truebp}

\np

\psfig{file=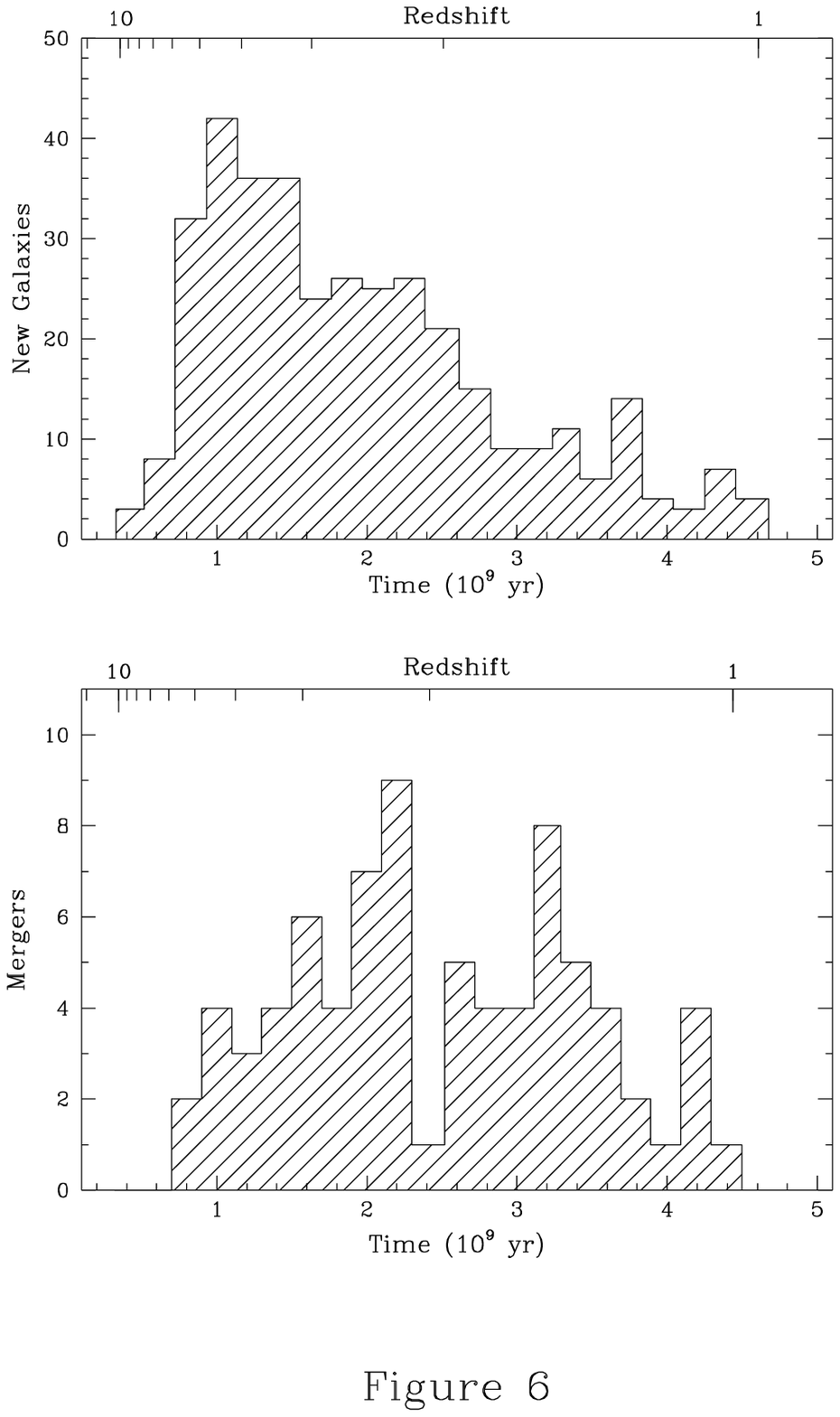,width=574truebp}

\np

\psfig{file=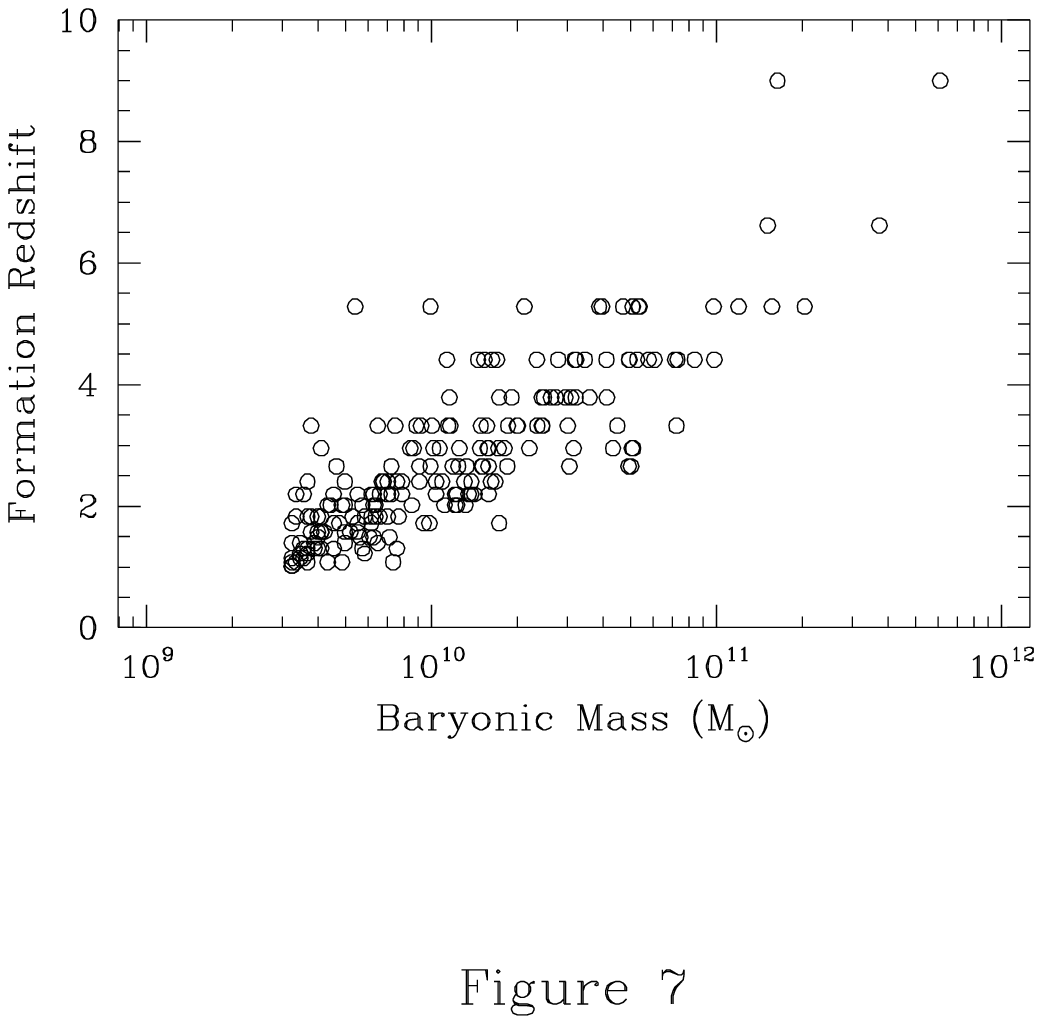,width=574truebp}

\np

\psfig{file=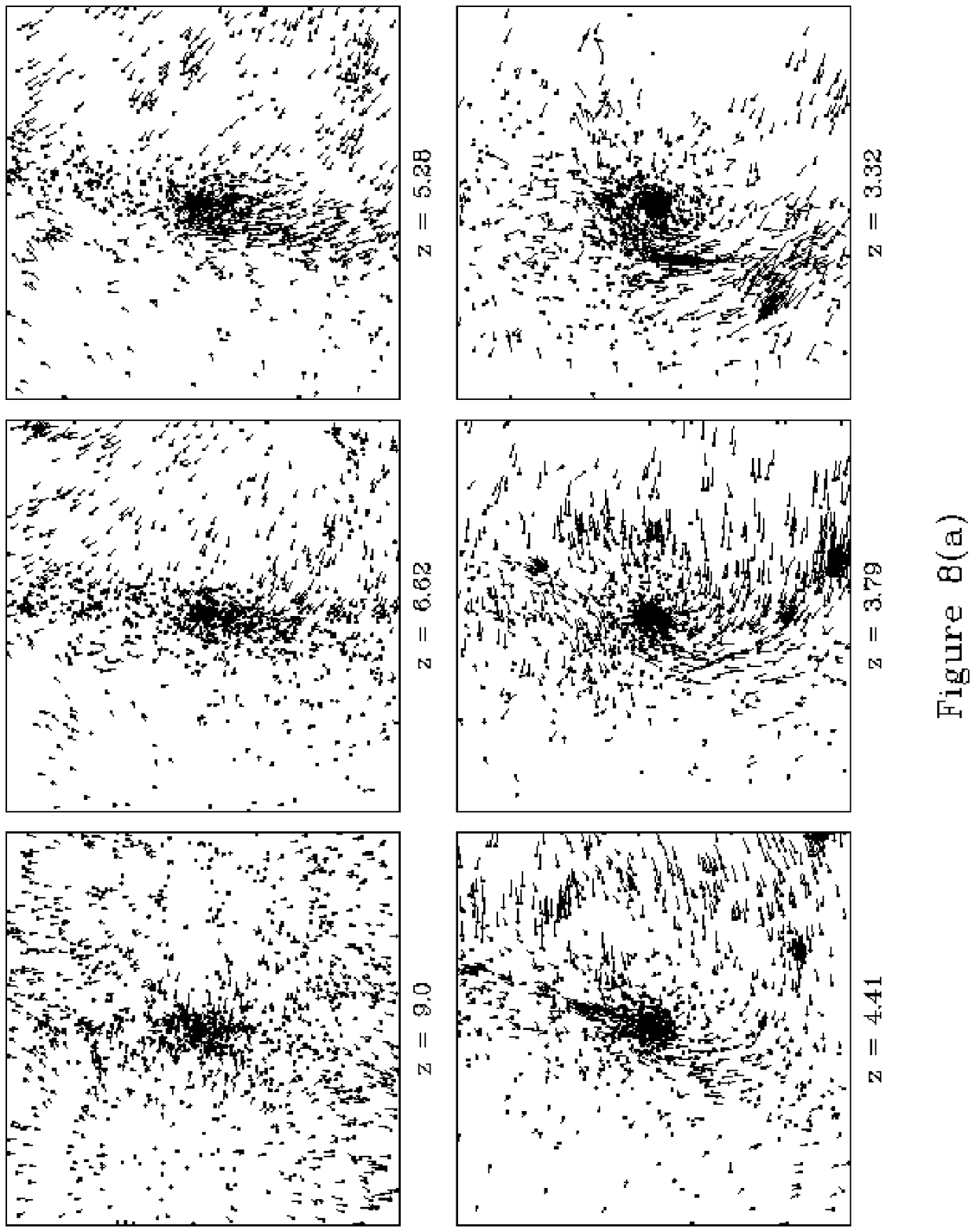,width=560truebp}

\np

\psfig{file=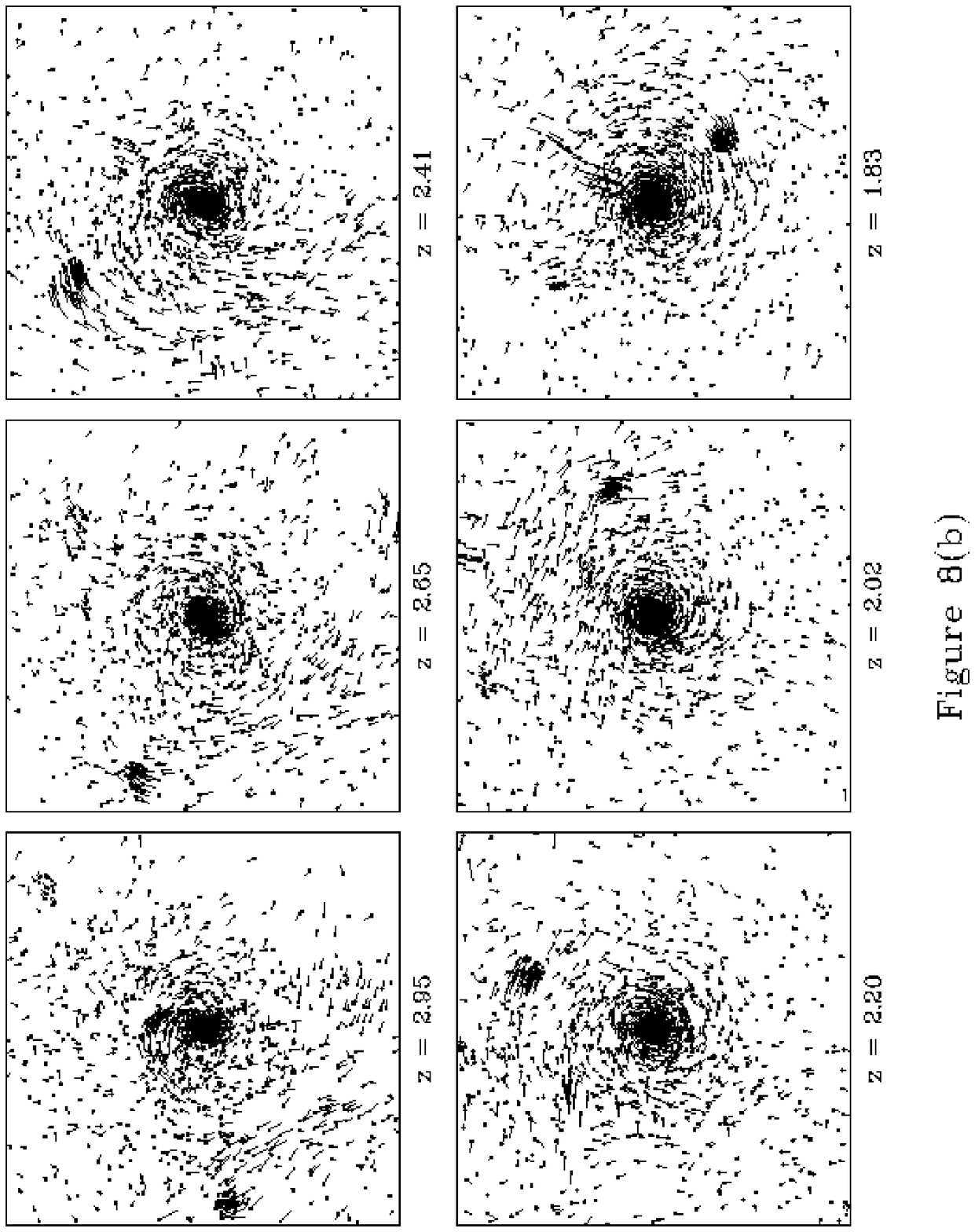,width=560truebp}

\np

\psfig{file=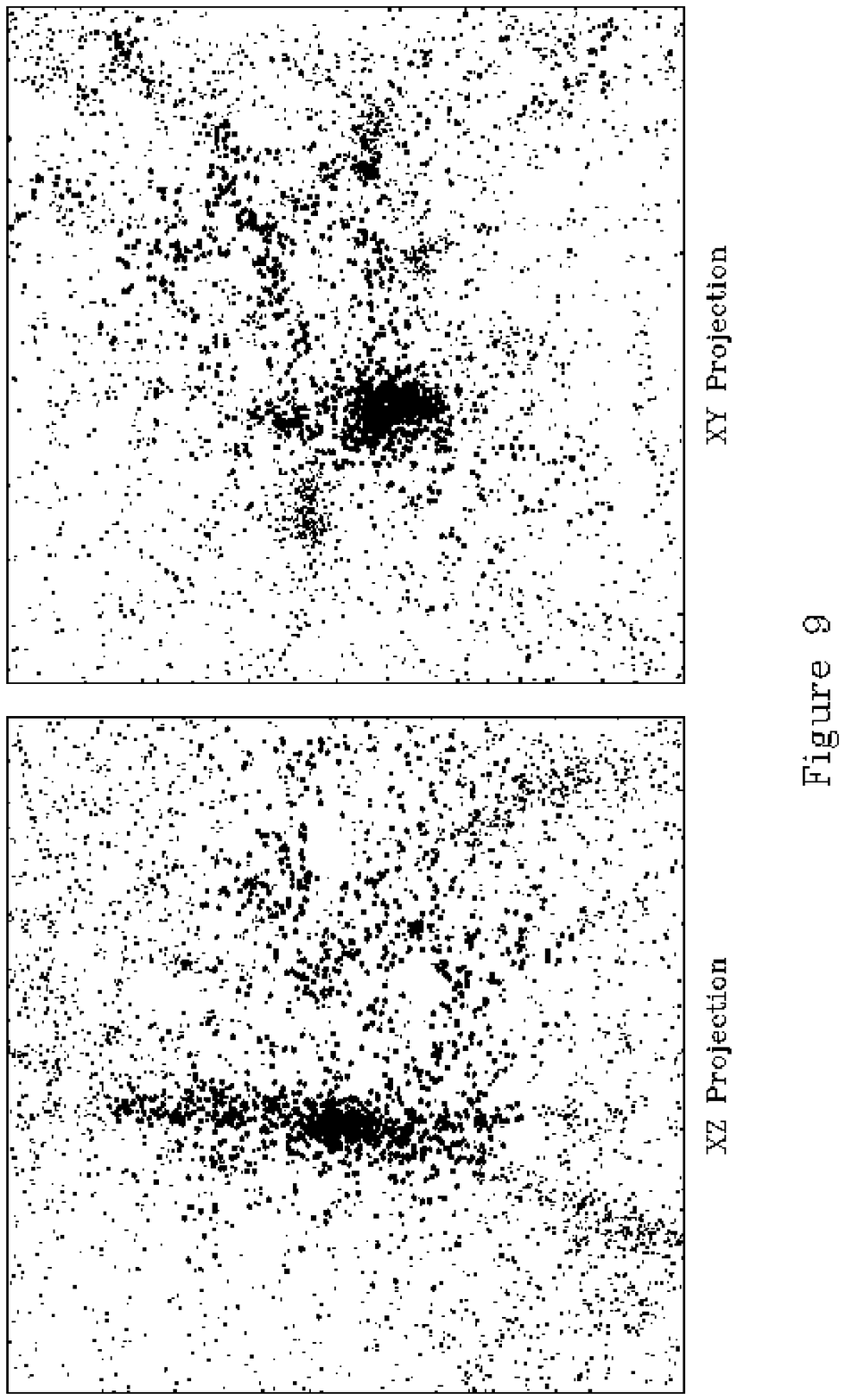,width=560truebp}

\np

\psfig{file=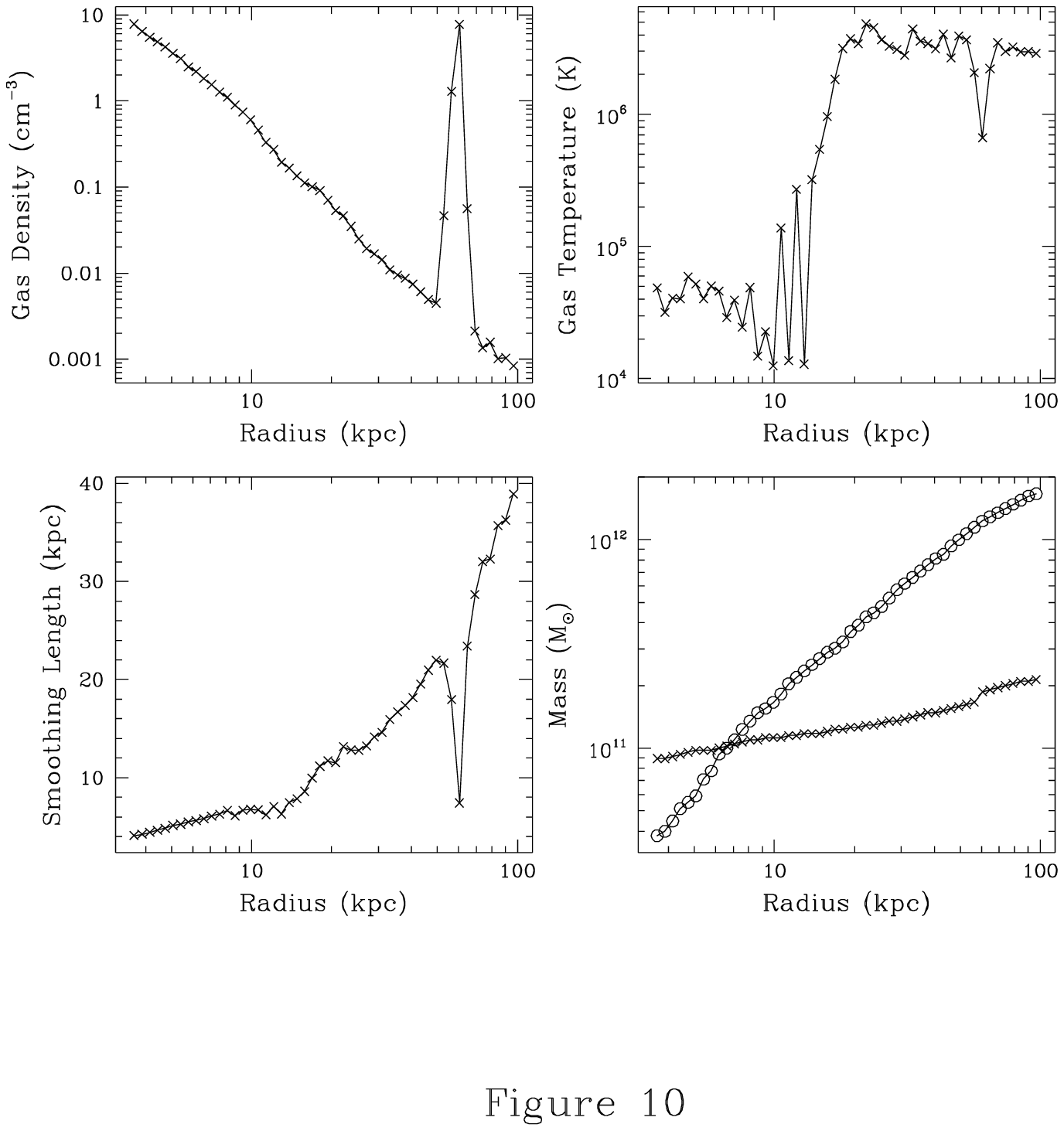,width=574truebp}

\np

\psfig{file=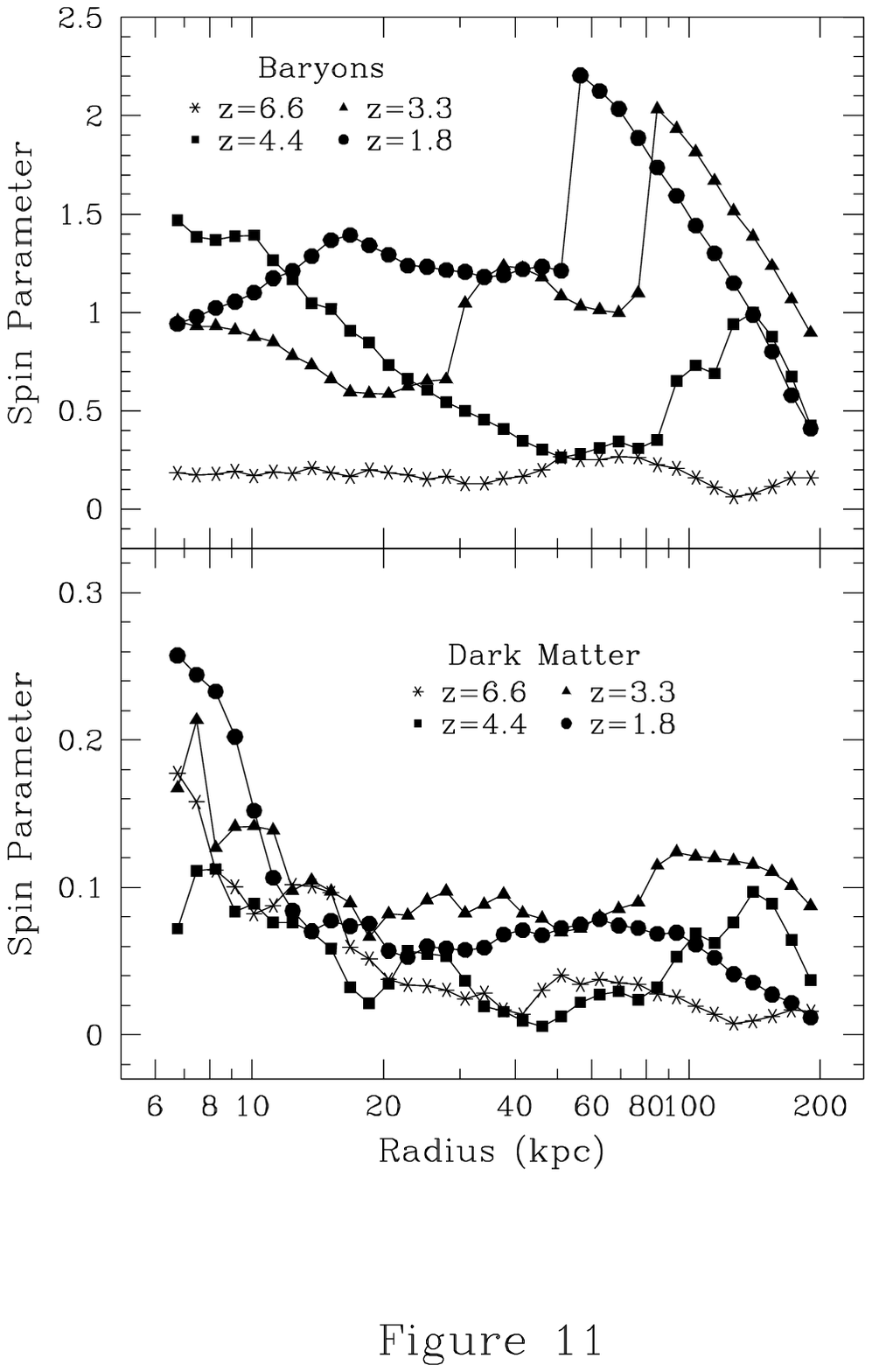,width=574truebp}

\np

\end